\documentclass[pdflatex,sn-mathphys-num,iicol]{sn-jnl}
\usepackage[english]{babel}
\usepackage{amsmath,amssymb,enumerate}
\usepackage{multicol}
\usepackage[tikz]{mdframed}
\usepackage{graphicx}

\usepackage[title]{appendix}%
\usepackage{xcolor}%
\usepackage{textcomp}%
\usepackage{manyfoot}%

\newcommand{\infixand}{\text{ and }}

\newcommand{\nobracket}{}
\newcommand{\textdots}{...}

\newcommand{\tmem}[1]{{\em #1\/}}
\newcommand{\tmmathbf}[1]{\ensuremath{\boldsymbol{#1}}}
\newcommand{\tmop}[1]{\ensuremath{\operatorname{#1}}}
\newcommand{\tmstrong}[1]{\textbf{#1}}
\newcommand{\tmtextit}[1]{\text{{\itshape{#1}}}}

\mdfsetup{linecolor=black,linewidth=0.5pt,skipabove=0.5em,skipbelow=0.5em,hidealllines=true,innerleftmargin=0pt,innerrightmargin=0pt,innertopmargin=0pt,innerbottommargin=0pt}
\newmdenv[topline=true,bottomline=true,innertopmargin=1ex,innerbottommargin=1ex]{tmbothlined}

\begin{document}

\title[Title]{Relativistic quantum mechanics and quantum field theory}

\author[]{\fnm{Urjit A.} \sur{Yajnik}}\email{yajnik@iitb.ac.in}

\affil[*]{\orgdiv{Department of Physics}, 
\orgname{Indian Institute of Technology Gandhinagar}, 
\orgaddress{ \city{Gandhinagar}, \postcode{382055}, \state{Gujarat}, \country{India}}}



\abstract{  Relativistic quantum mechanics can be considered to have begun with a search
  for wave equations corresponding to each intrinsic spin. However,
  relativistic quantum physics differs fundamentally from the non-relativistic
  wave mechanics. It requires a formalism allowing \ creation and destruction
  of particles.  This gets proper treatment only in a framework called
  quantum field theory. This article is a semi-historic account of the
  intriguing new features which emerge as a part of quantum field theory. Such
  a discussion is impossible without a basic presentation of the formalism
  itself. Hence some mathematics is included in finer print. The article is
  directed mostly to those familiar with essential classical mechanics and
  basic quantum mechanics, though I
  strive to provide a flavour of the subject to the keenly interested
  non-physics reader.}

\maketitle

\tableofcontents

  \section{Introduction}
  
  The fundamental way in which relativistic quantum mechanics differs from
  the non-relativistic wave mechanics is the possibility of creation and
  destruction of particles. The Sch{\"o}dinger wave function is normalised to
  unity; this fixes the number of quanta to one. In many-body systems this is
  generalised to fixing the norm to the total number of particles.
  However this framework cannot treat photons, which can be  emitted and
  absorbed by atoms and molecules. Similarly, high energy cosmic rays 
  showed that particle tracks changed from one type to another, that is 
  destroying one type and creating another type of particle.
  Thus a quantum theory was needed in which quanta could be
  created and destroyed consistent with special relativity.
  
  The route to such a theory was not easy, nor straight. One key development
  was the Dirac Equation which automatically included anti-particles. Dirac
  found an ingenious single particle interpretation for his equation, called
  ``hole theory''. But its more comprehensive interpretation remains
  many-particle. Another development was quantum theory of radiation. This was
  pioneered by the trio who pioneered quantum mechanics, Born, Heisenberg and
  Jordan. Their formalism introduced the creation and destruction operators
  for the photons. This was soon generalised to the case of massive charged particles
of spin $0$  as well as to the electron, of spin $\hbar/2$.
Interestingly the latter generalisation required  
  that the creation and destruction operators for electrons
  should obey {\tmem{anti-commutation}} relations rather than commutation
  relations. This is equivalent to the spin-$1 / 2$ particles obeying the
  Pauli exclusion principle. 
    However, proper interpretation of  these early formalisms required
  quantisation of space-time fields. This latter development happened through
  ``second quantisation'', which sometimes looked as if the wave function was
  being quantised. These developments are discussed in Sec. \ref{sec:qbf}.
      
  The next major development was Quantum Electrodynamics (QED), requiring
  renormalisation formalism. The language Richard Feynman developed for this theory is
  now common lore, the Feynman diagrams, the propagator, the anti-particles
  that propagate backwards in time etc. This fascinating period of developments is
  covered in Sec. \ref{sec:qed}. Over a decade later, these methods
  were put in a systematic perspective by Steven Weinberg, highlighting the role of
  the Poincar{\'e} group of symmetries in the theory of massive and massless
  particles. This is presented in Sec. \ref{sec:modernqft}. 
  
  The highpoint of relativistic quantum theory, i.e., quantum field theory,
  (QFT) was reached in the subtle prescription of renormalisability. First
  developed for QED, this method remained enigmatic at first. However, the two other
  fundamental forces, the weak and strong nuclear force could also be formulated 
  successfully as renormalisable quantum field theories by the early 1970's. 
  These theories are together called the Standard Model of elementary particles. The
  Higgs boson, an important member of the Standard Model was the last to be discovered
   at the Large Hadron Collider in 2012. Soon after the emergence of the Standard Model, 
   a physical  understanding of renormalisability appeared in the works of Kenneth Wilson and Joseph 
  Polchinski. We will not be able cover these developments here.
  
  Other than this mainstream development of the formalism, there were two key
  issues that became cornerstones of relativistic quantum theory. These two
  being, the spin-statistics theorem and the $C P T$ theorem. These are
  overarching principles that work as touchstones for much of QFT. Any new
  new particles and their interactions we try to introduce must obey these principles.
   Whether the principles always hold is an open question,
  especially the $C P T$ theorem. We shall discuss the significance of these
  two principles in Sec. \ref{sec:CPT}
  
  Overall, we may say, quantum field theory is the valid mature successor
  theory of the initial approaches towards relativistic version of quantum
  theory. QFT appears to minimally encompass everything that relativity
  mandates in quantum theory, a necessary and sufficient framework at the present 
  state of knowledge. QFT has met several spectacular successes. It is
  most effective in the ``perturbative regime'' where the interaction between particles can be
  treated as weak and localised effects in scattering experiments. Another approach where 
  QFT methods are used is called ``effective field theory'' approach. This
  is useful when the available energy can be considered limited and then these methods 
  can be used for all observables below that energy scale. 
  But beyond these expansion methods, QFT also yields tantalising information about  
  non-perturbative phenomena, without any restriction on the strength of
  coupling or energy scale. This usually requires what are called the functional methods, 
  a generalisation of the path integral method of Feynman. The functional
  methods have often involved topological considerations of extended
  field configurations in space-time. The one important problem not satisfactorily 
  addressed is that of relativistic bound states. The formalisms that have been developed 
  to address this have had limited success.
  
  Since the mid-1970's many path breaking proposals have been made to go beyond the 
  Standard Model, but they have not met with any success. In this context, string theory
  has been the most comprehensive and appealing new theory that incorporated
  the relativistic quantum principles and promised to be the "theory of everything".
  But the predictions of the theory have not been borne out by the observed Universe.
  Furthermore, it turns out that even such a comprehensive theory will also eventually need 
  a QFT description for its low energy interpretation. So in a sense, ``QFT rules".
  
  This article is an attempt to tell the story of these developments through
  anecdotes and major landmark results. For a more exhaustive history, the reader is 
  referred to \cite{SW-v1}, chapter 1. There are a number of excellent textbooks on QFT
  many of them online. However, one comprehensive reference that includes many a subtle calculation is \cite{IandZ}.
  
  \section{Peculiarities of fundamental particles}
  \label{sec:peculiarities}
  
  We begin with a few essential features of the physical system the theory is meant to describe. 
  
    Relativistic quantum theory of Dirac required the existence of
  anti-particles. This was the first novelty encountered. Anti-matter has since been a
  fascinating element in many a sci-fi narrative. Secondly, it naturally
  incorporated the spin of the electron of value $\hbar / 2$. What is unusual
  about spin is that it is unlike any rotation. It can not be obtained via the angular momentum formula
  $\tmmathbf{r}  \times \tmmathbf{p}$ with electron at the origin. This is because we have to treat the electron as
  a point particle and $\tmmathbf{r}$ is zero. Thus, although no classical sense of
  rotation is possible for an idealised point, we find that this is indeed
  something like rotation. The intrinsic spin adds on to the atomic orbital angular momentum of the electron. As of now we have
  no intuitive understanding of the origin of the spin of particles. But abstract mathematics seems to bring in spin very
  naturally from trying to think of the particles as manifesting or ``representing'' the symmetries of special relativity, the so called Poincar{\'e} group of symmetries.
  
  There are at least two other features learnt from experiments which deserve
  to be highlighted. One can be called ``identity crisis''. In this, a set of
  particles can ``oscillate'' back and forth between related particle species.
  For example a beam of electron neutrinos produced from a nuclear reactor,
  when observed some distance away, will reveal the presence of muon neutrinos
  and also tau neutrinos, conserving the total number of neutrinos. This phenomenon 
  occurs in a few other systems of fundamental particles as well, which contain the strange quark and the bottom quark. At the heart of this 
  curiosity is also an unexplained fact of nature. This is that there are three families 
  of the fundamental fermions, both among the quarks and the leptons. For example,
  there are fundamental particles called the muons and the tau leptons, which have 
  all the quantum numbers and spin the same as the electron, but their mass values are widely different.
    
  The second
  intriguing point may be called ``hidden sectors''. This means that particles
  need not necessarily ``interact''. Unlike in the physics of normal
  experience, where two bodies necessarily recoil when pushed together, here
  there can be particles that simply never ``see'' each other and can ``pass
  through''. In the heyday of grand unification, among ideas based on supergravity, a hidden 
  sector was a standard ingredient of the theory. The possibility of hidden sectors raises 
  an intriguing 
  question for Dark Matter. As we know, Dark Matter is needed in substantially large proportion
  to explain the observed expansion rate of the Universe. We
  therefore hope that it will show effects in our detectors as well. But it
  may well be that the only interaction it has with usual matter is
  gravitational. So although streaming through the apparatus, its interaction could 
  be far too weak to be detected in individual scattering events.
  
  \section{Dirac equation and hole theory}\label{sec:diractheory}
  
  We begin now with first of the breakthroughs of relativistic quantum theory. 
  Historically, the Dirac equation for the electron followed as a sequel to
  Schr{\"o}dinger's non-relativistic equation. However, the 
  latter was the prequel, since special relativity is the foundational principle of physics, while non-relativistic physics is an approximation. 
  
  Schr{\"o}dinger proposed his well known equation
  \begin{equation}
    i \hbar \frac{\partial}{\partial t} \psi = - \frac{\hbar^2}{2 m} \nabla^2
    \psi + V \psi
  \end{equation}
  A simple rule that explains this equation is that it is the quantum version
  of the Hamiltonian (same as energy for the present purpose), expression
  \[ H = \frac{\tmmathbf{p}^2}{2 m} + V \]
  This is now interpreted as an operator equation with substitution by
  differential operators
  \begin{equation}
    H \rightarrow i \hbar \frac{\partial}{\partial t}, \qquad \tmmathbf{p}
    \rightarrow - i \hbar \nabla \label{eq:Qrules}
  \end{equation}
  Furthermore, electromagnetism can be introduced using the rules,
  \begin{equation}
    H \rightarrow H - q \phi ; \qquad \tmmathbf{p} \rightarrow \tmmathbf{p}-
    \frac{q}{c} \tmmathbf{A}
    \label{eq:minimalcoupling}
  \end{equation}
  where $\phi$ and $\tmmathbf{A}$ represent the scalar and the vector
  potentials. This recipe reproduces the Lorentz force for a charged particle
  correctly. Remarkably, Schr{\"o}dinger obtained the main values of Hydrogen atom levels correctly  by
  solving this equation because in Hydrogen atom specifically, the electron motion is non-relativistic.\
  
  An obvious extension of this procedure is to implement the relativistic  relation
  \begin{equation}
    E^2 =\tmmathbf{p}^2 c^2 + m^2 c^4 \label{eq:relativisticrelation}
  \end{equation}
  with the same rules for going to operators, and obtaining a relativistic wave equation. Apparently Schr{\"o}dinger
  did try this for the Hydrogen atom. But the differences of his energy levels
  did not match the observed energies of spectral lines using the famous
  quantum principle
  \begin{equation}
    E_n - E_m = h \nu_{m n} = \hbar \omega_{m n}
  \end{equation}
  where $E_n$, $E_m$ are the energy levels, $\nu_{m n}$ is the frequency and $\omega_{m n}$ is the angular frequency of the emitted radiation, with $\hbar=h/2\pi$. The reason why he did not get these
  correctly is because at that time it was not known that the net angular
  momentum of the electron in the Hydrogen atom is half-integral due to its
  intrinsic spin.
  
  The relativistic wave equation known as the  Klein-Gordon equation
  \begin{equation}
    - \hbar^2 \frac{\partial^2}{\partial t^2} \phi + \hbar^2 c^2 \nabla^2 \phi
    - m^2 c^4 \phi = 0 \label{eq:KGeqn}
  \end{equation}
  is obtained by introducing the operators of \eqref{eq:Qrules} in the relation
  Eq. \eqref{eq:relativisticrelation}.
  Upon its first introduction, a
  major deficiency was observed, namely, its interpretation as a probability
  amplitude wave was inconsistent. The conserved density it provided was
  indefinite in sign, i.e., it could also give a negative answer for
  probability, which is nonsense. Hence the conserved quantity could not
  represent probability density. This is in contrast to $\psi^{\dag} \psi$
  from the non-relativistic equation which is positive definite (positive or
  zero). Actually the problem was of interpretation, not of substantial
  validity. But historically this resolution had to await other developments.
  
  In 1929 Dirac was intensely engaged with the problem of reconciling Special
  Relativity with the quantum rules Eq. (\ref{eq:Qrules}). He also knew that if
  the equation were first order in time derivative, it would produce a
  positive definite probability density. The energy momentum relation
  (\ref{eq:relativisticrelation}) could be rendered $E  = \pm (\tmmathbf{p}^2 c^2
  + m^2 c^4)^{1 / 2}, $ in which case the wave equation would become first order in time, 
  \begin{equation}
    i \hbar \frac{\partial}{\partial t} \phi = \pm (- \hbar^2 c^2 \nabla^2 +
    m^2 c^4)^{1 / 2} \phi \label{sqrooteqn}
  \end{equation}
  But what to make of the
  square-root of the operators on the right? At this point Dirac was visiting
  Niels Bohr at his institute in Copenhagen. Bohr recalls being perplexed to
  hear that Dirac was trying to take the square root of a matrix. Bohr would
  have been even more perplexed, had he known that Dirac was effectively
  trying to take the square-root of the identity matrix \cite{Gottfried:2010kn}.
  
  Dirac searched for coefficients $\alpha_i$, and $\beta$, \ such that the operator
  \[ i \hbar c \left\{ \alpha_1 \frac{\partial}{\partial x} + \alpha_2
     \frac{\partial}{\partial y} + \alpha_3 \frac{\partial}{\partial z} +
     \beta m \right\} \]
  when squared would produce the operator under the square-root sign of Eq.
  (\ref{sqrooteqn}). He indeed found the needed coefficients, except that they
  could not be numbers but were matrices. In modern usage, instead of
  $\alpha_i$, and $\beta$ one defines matrices $\gamma^{\mu}$, with
  $\mu = 0, 1, 2, 3$, which tallies with the time and space
  components of Lorentz 4-vectors of Special Relativity.
  
  ***

 {\scriptsize Let us interpret what Dirac found. First we rewrite Dirac's proposal, which
  became the Dirac equation, in modern notation as
  \begin{equation}
    i \left\{ \gamma^{0} \frac{\partial}{c\partial t} + \gamma^{1} \frac{\partial}{\partial x} + \gamma^{2} \frac{\partial}{\partial y}
    + \gamma^{3} \frac{\partial}{\partial z} \right\}\psi = m \psi
\label{eq:diracequation}
  \end{equation}
  What he looked for was that applying the left hand side operator twice, while
  obtaining $m^2$ on the right, one should recover the Klein-Gordon equation
  (\ref{eq:KGeqn}). 
  In order to reproduce the Klein-Gordon equation, the $\gamma$-matrices have
  to satisfy
  \begin{equation}
    \gamma_{\mu} \gamma_{\nu} + \gamma_{\nu} \gamma_{\mu} = 2 \mathbb{I}
    \eta_{\mu \nu} \qquad \mathrm{for} \quad \mu, \nu =0,1,2,3
     \label{Cliffordalg}
  \end{equation}
  with $\mathbb{I}$ the $4 \times 4$ identity matrix. 
  The symbols $\eta_{\mu \nu}$ are such that $\eta_{i j}=-\delta_{i j}$ for $i = j = 1, 2, 3$, while 
  the only other non-zero element is $\eta_{00} = + 1$. As such,
  a non-zero $+ 1$ or $-1$ occurs on the right only when $\mu = \nu$ on the left hand side, the
  relevant matrix being squared. In this sense the $\gamma$-matrices are a
  square-root of the ``identity matrix", actually the metric tensor of the Minkowski space.}

  ***

  \subsection{Hole theory}
  
  Dirac discovered that the required size of the $\gamma$-matrices is $4
  \times 4$, and the spinor wave function $\psi$ had to be a $4$ component
  column vector, ie., 4 independent complex wave functions. Dirac could
  identify 2 of the components as spin up and spin down states of the
  electron, but the other two corresponded to spin up and spin down with 
  negative energies. It seemed
  like the ghost haunting the relativistic version was going to resurface in
  some form or the other. On the face of it, this was a catastrophe, since if
  negative energy levels were available, particles could jump spontaneously to
  lower energy states, but there were levels with indefinitely large negative
  values and the total energy of electronic matter could become negative
  infinity! 
  
\begin{figure}[htb]
        \centering
        \includegraphics[width=\linewidth]{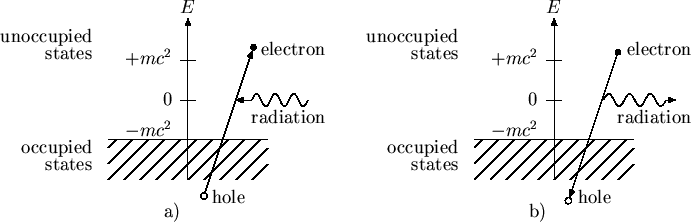}
        \caption{Dirac's proposal for hole theory. Empty levels in the otherwise completely filled "negative energy sea" appear to us to be positively charged electrons of positive energy. Left panel shows a photon imparting its energy to a negative energy electron, lifting it to a positive energy level. This appears to us as electron-positron pair production. The right panel shows  a normal electron jumping to fill a hole in the negative energy sea, emitting a photon in the process. This appears to us as pair annihilation. Graphic courtesy \cite{Gingrich}.}
        \label{fig:holetheory}
    \end{figure}

  At this point Dirac assumed
  that all the negative energy states were fully occupied. Then invoking the Pauli exclusion
  principle prevented jumps of any other electrons into those levels. The
  negative energy solutions also carry rest mass $-m$, negative of the mass of
  the electron. Dirac then interpreted the \tmtextit{absence} of negative
  mass, negative energy electron from the otherwise fully packed negative energy sea, as the {\tmem{presence}} of a positive mass
  positive energy electron and of the opposite charge. The only particle of
  exact same value but opposite sign of the electron charge known at that time
  was the proton.
  Although their masses were in the ratio 1:2000, Dirac boldly identified the
  proton as the anti-electron.  Fig. \ref{fig:holetheory} illustrates some concepts of hole theory.
  
  \subsection{The $g$ factor}
  \label{subsec:gfactor}
  
  The electron has a charge $(- e)$ and also spin $\hbar / 2$. A classical
  charged sphere set spinning would possess a magnetic moment as an Amp{\`e}re
  effect, viz., current carrying loops generate a magnetic field. Of course
  the value of the generated magnetic field would depend upon the distribution
  of the charges in, and the angular speed of the sphere.  A general way of representing this is to say,
  \begin{equation}
\tmmathbf{m}= g\tmmathbf{L}
\label{eq:gfactor}
  \end{equation} 
    that is, the magnetic moment vector $\tmmathbf{m}$ of that spinning object would be
  aligned with the angular momentum vector  $\tmmathbf{L}$ of the latter, with some proportionality constant $g$. For example an atom has many electrons in various orbitals. The
  combined magnetic moment that can arise from this substructure would be
 proportional to the net angular momentum of the atom. In spectroscopy this proportionality is known as the Land{\'e} $g$-factor.
  
  But what about the electron itself? It is a point particle so there is no
  sense to its ``spinning''. And its $g -$ factor had a measured value 2! If
  one created some model of the electron as an extremely compact charge
  distribution spinning in the conventional sense, the $g$-factor cannot reach
  such a large value. The reassuring feature of Dirac's rather exotic ( at
  that time) formalism was that it correctly predicted $g$ to be 2.
  
  \subsection{Limitations of the one-particle interpretation}
  
  With the hindsight of history we know that Dirac had
  invented (indeed invented, not just ``discovered'') the right
  equation born of wrong motivations. Indeed, Dirac could obtain the positive
  definite probability density, but that is not how the equation was going to
  be used eventually in elementary particle physics.  However, the hole 
  concept as elaborated by Dirac serves well in band
  structure of solids.
  
  Dirac's bold conjecture about the existence of oppositely charged electrons
  was borne out, not as the proton, but by the appearance of positrons in
  cosmic rays. In 1932 C. D. Anderson, apparently unaware of Dirac theory,  discovered a charged particle track of
  the same mass as the electron but turning oppositely in an applied magnetic
  field. He then also demonstrated the existence of such a particle in a
  laboratory experiment. He received the Nobel Prize in 1936.

  With hindsight we also know that electron and positron, indeed any pair 
  of particle-anti-particle are on exactly equal footing in the fundamental theory. 
  Therefore a view of positrons as reinterpreted holes is artificial. Further, zero spin and integer
  spin particles which do not obey the exclusion principle also have negative energy solutions.
  So we cannot propose a fully filled Dirac sea for them. Thus the negative energy solutions 
  must always be reinterpreted as normal particles of opposite charge. This raises a 
  fundamental question of cosmology. If both matter and anti-matter are on the same footing,
  why does the Universe today have only  protons and electrons, but no natural abundance of
  anti-protons or positrons. These latter particles are seen frequently in laboratory experiments, 
  but they just disappeared from the universe as we see it today. In modern 
  particle physics we seek solution to this puzzle in dynamical processes in early the big bang universe. Majorana neutrinos which are discussed later, are expected to play an important role in such a resolution.
  
    \section{Quantisation of bosons and fermions}
  \label{sec:qbf}
 
 A key new element of the relativistic theory is {\tmem{creation}} and
  {\tmem{destruction}} of quanta. This comprehensive understanding was yet emerging 
  in the early years of quantum mechanics. But for photons it was already well
  known, as emissions and absorptions by atoms and molecules. In the second of the two famous 
  papers of Born Heisenberg and Jordan \cite{Born:1926uzf},
  they proposed a way to implement creation and destruction operators for the case of photons.
  
  A more comprehensive framework soon
  emerged, developing quantisation of different particle species more generally. This was 
  due to a variety of new particles being discovered, either in the laboratory or the cosmic rays.
  A prime new discovery was that of the neutron as an important and ever present ingredient
  of nuclei. The discovery of the neutron had surprisingly taken very long, 
  till 1932. But it immediately became obvious that a new class of force would have 
  to provide the very strong binding between the proton and the electromagnetically 
  neutral neutron.  Beta decay could now be described as a transmutation of a nuclear 
  neutron into a proton with the emission of an electron. Further, new particles, the 
  "mesons"  were being discovered in cosmic rays. It was clear that a general quantum 
  theory for production and destruction of such quanta needed to be in place.

  \subsection{Radiation theory}\label{sec:radiation}
 
  The work of Born Heisenberg and Jordan introduced a very durable recipe into
  quantum theory. There is an interesting analogy between the quantum harmonic oscillator and
  the state containing a number of quanta. The spacing between the energy
  eigenvalues of the harmonic oscillator is constant, in usual notation, $h\nu$=$\hbar \omega$. 
  Now a state of $N$ photons of the same frequency will have energy $N h\nu$, as if it 
  were the $N$th excited state of the harmonic oscillator of frequency $\nu$. 
  Thus, such states can be treated very
  efficiently using the algebra of the quantised harmonic oscillator. We need
  to introduce one oscillator for every possible frequency value. The usual notation uses 
  wave number \tmmathbf{k=\omega/c}, thus one oscillator needs to be introduced
  for each value of $\tmmathbf{k}$. 
  Similar approaches to photon quantisation were adopted by Dirac, Fock and others.
 
 ***
  
  {\scriptsize The harmonic oscillator in one space dimension is described by the
  Hamiltonian
  \begin{equation}
  H = \frac{1}{2 m} p^2 + \frac{1}{2} m \omega^2 x^2 
  \label{eq:HO}
  \end{equation} 
  where we have chosen the conventional force constant $k \equiv m \omega^2$,
  and the significance of $\omega$ as the angular frequency of oscillations
  emerges after the problem is solved. This expression is of the form $X^2 +
  Y^2$ which can be written as $Z Z^{\ast}$ where $Z = X + i Y$ and $Z^{\ast}$
  is the complex conjugate. Similarly here we define
  \[ a = \sqrt{\frac{m \omega}{2 \hbar}} x + i \sqrt{\frac{1}{2 m \omega
     \hbar}} p \]
  so that
  \[ H = \hbar \omega a a^{\ast} \]
  Now in quantum theory $p$ and $x$ become operators, obeying
  \[ \left[ x, \, p \right] = i \hbar \]
  Now suppose we re-express $x$ and $p$ in terms of the $a$ and $a^{\ast}$, but 
  denoting the operator for $a^{\ast}$ as $a^{\dag}$. The the above condition becomes
  \[ \left[ a, \, a^{\dag} \right] = 1 \]
  The corresponding quantised Hamiltonian has the expression
  \[
  H = \hbar \omega \{ a^{\dag} a + \frac{1}{2}\}
  \]
  And we need to introduce one harmonic oscillator degree of freedom for each desired
  unit of energy interval $\Delta E_{\tmmathbf{k}} = \hbar \omega_{\tmmathbf{k}}$.
  Note that the extra $\frac{1}{2}$ in the quantum Hamiltonian is a perfectly legitimate ground state energy for
  a harmonic oscillator. However, in this improvised recipe, it is an
  embarrassment as one gets an infinite tower of vacuum energy contributions 
  from each value of the frequency in the total Hamiltonian.
  But one has to remember that proposing a product operator from classical analogy as in Eq. \eqref{eq:HO} is only a convenience. Ideally there is an independent quantum observable $H$.
  We give up this independence and retain our convenience, by ``normal ordering"
  the above expression in the case of a field system. The details can be found in QFT textbooks, including \cite{IandZ}.
}

***
  \subsection{Quantisation of charged bosons}
  Klein and Jordan proposed a reinterpretation of the Klein-Gordon dilemma.
  The problem of Klein-Gordon equation was that the conserved current which
  has the interpretation of probability density current is in the danger of
  becoming negative and hence unphysical. In the intervening years, cosmic
  rays had shown the existence of new particles called the pi ($\pi$) mesons.
  By observing the direction in which their trajectories bend in applied
  magnetic field one can find that there are $\pi$-mesons of both charges,
  $\pm1$ but of exactly the same mass.
  
  Klein and Jordan proposed that the complex valued scalar field can be
  considered to be a description of the $\pi$-mesons. The current of
  indefinite sign can then be interpreted as being the current of electric
  charge, which can have both signs. They placed the quantum conditions not on
  the position coordinate of the pions, nor expect pion number to be
  conserved. They instead
  imposed the quantisation conditions on the field $\phi$ itself, considered as
  a field valued dynamical variable. In other words they proposed  the conditions 
  \begin{equation}
    [\phi (\tmmathbf{x}, t), \pi (\tmmathbf{x}', t)] = i \hbar \delta^3
    (\tmmathbf{x}-\tmmathbf{x}'), \label{eq:canquantscalar}
  \end{equation}
  and the related  conditions
  \begin{equation}
    \left[ \phi (\tmmathbf{x}, t), \, \phi (\tmmathbf{x}', t) \right] = [\pi
    (\tmmathbf{x}, t), \pi (\tmmathbf{x}', t)] = 0 
    \label{eq:canquantscalar2}
  \end{equation}
  where $\pi$ is the canonically conjugate momentum field as per Hamiltonian mechanics.
  These additional conditions assert the independence of the fields at different space points
  $\tmmathbf{x},$ $\tmmathbf{x}'$ at the same time $t$.
  
  These approaches to the quantisation of photons and spin-$0$ particles showed that the 
  ``indistinguishability'' of photon quanta as first formalised  by S. N. Bose is 
  captured correctly if we impose the conditions
  \begin{eqnarray}
    {}[a_{\tmmathbf{k}}, a^{\dag}_{\tmmathbf{k}'}] \equiv & a_{\tmmathbf{k}}
    a^{\dag}_{\tmmathbf{k}'} - a^{\dag}_{\tmmathbf{k}'} a_{\tmmathbf{k}} = &
    \delta_{\tmmathbf{k}, \tmmathbf{k}'} \nonumber\\
    {}[a_{\tmmathbf{k}}, a_{\tmmathbf{k}'}] = & [a^{\dag}_{\tmmathbf{k}},
    a^{\dag}_{\tmmathbf{k}'}] = & 0  \label{eq:fockoscillators}
  \end{eqnarray}
  All the particles that have intrinsic spin $0$ or $1$ in the units of $\hbar$, and all other integer spin values in general,  obey Bose 
enumeration. This is often mistakenly called Bose statistics, because of the use
of the counting in statistical mechanics. For all such quanta, the quantisation
conditions have to be as in Eq. (\ref{eq:fockoscillators}) above.  
  
  We thus construct a grand Hilbert space, by stitching together one copy of the harmonic oscillator  Hilbert space for every
  value of the wave-number vector $\tmmathbf{k}=\tmmathbf{p}/ \hbar$. And the
  states are interconnected by the operators that are the destruction operators $a_{\tmmathbf{k}}$ and creation operators
  $a_{\tmmathbf{k}}^{\dag}$. This algebra for its completeness requires a
  grand {\tmem{vacuum state}}, the state of no quanta at all. Including this
  state is in a sense as important as the zero symbol $0$ introduced in
  arithmetic in ancient India. The vacuum state is denoted $| 0 \rangle
  \nobracket$, or sometimes by $| \Omega \rangle \nobracket$. The vacuum is
  an important concept, as we shall see later in the section \ref{sec:qed} on Quantum Electrodynamics (QED).

Treating the scalar field itself as a configuration space variable also appeared in
  the works of Heisenberg and and Pauli, who had also realised the general applicability of
  the field quantisation procedure beyond the photons. 
 It was argued by Bohr and Rosenfeld that the quantum conditions can be interpreted as
 relativistically correct inobservability of influences  of field variables outside the light cone.
  With this framework, the probability interpretation of a wave field and  the amusement 
  of the celebrated Einstein-Podolsky-Rosen paradox do not arise any more. Probability
  interpretation now shifts to the probability amplitude for the destruction
  and creation of quanta of specific wave number.

  If $\phi$ were the wave function entering the Klein-Gordon equation, then
  the quantum conditions \eqref{eq:canquantscalar} and \eqref{eq:canquantscalar2}  
  amount to quantising the wave function. But wave function itself arose when
  quantising a single particle. This has
  led to the nomenclature second quantisation for this prescription of QFT.
\\

***

{\scriptsize
  {\flushleft{\bf The origin of the terminology ``second quantisation''}\cite{Dirac}}
  
  Suppose we have two equivalent bases for a single particle description,
  one labeled by the values of an
  observable $\alpha$ and another labeled by the values of an observable
  $\beta$. In quantum mechanics the  basis vectors would be related by a unitary
  transformation
  \[ | \nobracket \alpha_i \rangle = \sum_{\beta_j} U_{\alpha_i \beta_j} |
     \beta_j \rangle \nobracket \]
  Any operators in the system are expected to transform as
  \[ \mathcal{O}_{\alpha_i \alpha_k} = \sum_{\beta_l \beta_r} U_{\alpha_{i
     \beta_l}} \mathcal{O}_{\beta_l \beta_r} U_{\beta_r \alpha_k}^{\dag} \]
  But the destruction and creation operators have a different fate. Note that
  we can create the states listed above by
  \[ | \nobracket \alpha_i \rangle = a^{\dag}_{\alpha_i} \left| 0 \rangle
     \infixand \right| \beta_i \rangle = b^{\dag}_{\beta_j} | 0 \rangle
     \nobracket \]
  Since this is true for arbitrary states $| \nobracket \alpha_i \rangle$ and
  $| \nobracket \beta_i \rangle,$ the equation above implies that
  \[ a^{\dag}_{\alpha_i} = \sum_{\beta_j} U_{\alpha_i \beta_j}
     b^{\dag}_{\beta_j} \]
  So these operators have the apparent transformation property of state
  vectors, not operators. It appears that it is the wave function -- a Hilbert space vector
  -- which is being quantised. This arises due to obtaining the full many-particle Hilbert space using the single-particle Hilbert spaces. Of course the fully quantised theory implements
  the transformation between  $A$ and $B$, being the the many-particle  observables
  corresponding to  $\alpha$ and $\beta$
  through a suitable similarity transformation of the
  field operator expressions.
  }
  \\
  ***
  
  This constructive picture of the many-particle Hilbert space relies on the 
  wave-number, or the momentum values. It is then also possible to make a transformation from the wave number
  degrees of freedom $\tmmathbf{k}$ to space degrees of freedom
  $\tmmathbf{x}$. Ignoring the polarisation degree of freedom in photons for
  the moment, we go from operators $a_{\tmmathbf{k}} e^{- i
  \omega_{\tmmathbf{k}} t}$ which includes their time dependence, to
  space-time operators $\phi (\tmmathbf{x}, t)$ via the Fourier transform. 
  In the single oscillator language this is just the change back from $a$-$a^{\dag}$ to $x, \, p$
  variables.  This recipe of constructing the mathematics of many particle systems is
  called quantum field theory or QFT \cite{Dirac}.
  
  In retrospect, QFT is the only
  correct quantisation of relativistic systems. Neither the Klein-Gordon nor
  the Dirac equation has consistent interpretation as a single particle
  wave function. They make consistent sense only as QFT equations satisfied
  by quantised field operators. 
  
  
  \subsection{Quantisation of the electromagnetic field}
  
  The radiation field can now be more systematically quantised by using the
  canonical procedure. This was pioneered by Proca, Dirac and Fermi.  However, the progress along these
  lines is much more subtle because the electromagnetic potentials are subject
  to gauge transformations and therefore not unique. Retaining a mass term for the photon as adopted by Proca and letting it go to zero at the end of the calculations alleviates some of the problems. If we impose gauge
  conditions to fix the remaining freedom, the framework may not remain Lorentz covariant.
  These were the issues remedied by proposals of Suraj N. Gupta and K. Bleuler, eventually overcome in the Quantum Electrodynamics program pioneered by
  Schwinger and Feynman which we take up in the section on QED. 
  
***

{\scriptsize 
Maxwell's equations are equivalent to the lagrangian,
  \begin{eqnarray*}
    L & = & \int d^3 x \left( - \frac{1}{4} F_{\mu \nu} F^{\mu \nu} \right)\\
    & = & \int d^3 x (| \tmmathbf{E} |^2 - | \tmmathbf{B} |^2)
  \end{eqnarray*}
  This can be recast in a form more amenable to methods of classical mechanics by a few tricks. One is to use gauge potentials \ $\phi (\tmmathbf{x}, t)$ and $\tmmathbf{A}  (\tmmathbf{x}, t)$ instead of the fields
    $\tmmathbf{E}$ and $\tmmathbf{B}$. This results in the lagrangian containing conventional kinetic
  energy terms for the potentials. One can then hope to determine the Hamiltonian 
  in terms of the canonical field variables. Further, one imposes the Lorenz gauge,
  \[
  \frac{1}{c}\frac{\partial \tmmathbf{A}}{\partial t} + \nabla\cdot \tmmathbf{A} = 0
  \]
  One can also supply a mass term as Proca proposed. Then the equations look just like the Klein-Gordon equation \eqref{eq:KGeqn} and one can hope to use the same methods to quantise this system. However, the Lorenz gauge condition is difficult to interpret as an operator equation, which is where more salient procedure due to Gupta and Bleuler needed to be adopted.
}

  ***
  \subsection{Quantisation of fermions}
  
  The quantisation conditions (\ref{eq:canquantscalar}) and
  (\ref{eq:canquantscalar2}) are equivalent to the Born-Heisenberg-Jordan proposal for harmonic
  oscillator like operators $a_{ \tmmathbf{k}}$, $a^{\dag}_{ \tmmathbf{k}}$
  obeying conditions (\ref{eq:fockoscillators}). The commutation conditions allow
  construction of states with arbitrary number of identical bosons by acting
  upon the vacuum state. Such states are correctly symmetrised as per the
  principle of quantum indistinguishability, ie, the Bose enumeration of states. Now 
  when we  have electrons, they
  are also of an assembly of particles of identical mass values, and we expect to introduce similar
  creation operators for each momentum or wave number. However, here we 
  know we have the Pauli exclusion
  principle, i.e., no two electrons of identical quantum numbers can occupy
  the same state. More generally, it was realised that even when the quanta
  have different momenta, the correct quantum state is anti-symmetric under the
  exchange of the quanta. Indeed it is well known that if we have a spin $1 /
  2$ particle, a full circle rotation by $2 \pi$ brings it to negative of
  itself! Only after full circle rotation twice, or $4 \pi$ rotation does it
  return to itself. It was thus realised that for generating the states
  of quanta of $1 / 2$-integer spin, the conditions should be
  \begin{eqnarray}
    \{ a_{\tmmathbf{k}}, a^{\dag}_{\tmmathbf{k}'} \} \equiv & a_{\tmmathbf{k}}
    a^{\dag}_{\tmmathbf{k}'} + a^{\dag}_{\tmmathbf{k}'} a_{\tmmathbf{k}} = &
    \delta_{\tmmathbf{k}, \tmmathbf{k}'} \nonumber\\
    \{ a_{\tmmathbf{k}}, a_{\tmmathbf{k}'} \} = & \{ a^{\dag}_{\tmmathbf{k}},
    a^{\dag}_{\tmmathbf{k}'} \} = & 0  \label{eq:paulioscillators}
  \end{eqnarray}
  The curly brackets are called {\tmem{anti-commutators}}, due to the change
  of the relative sign. This simple modification of the quantum conditions
  automatically implies the required anti-symmetry and the Pauli exclusion
  principle. If the quantum number labels {\tmstrong{k}} are the same on two
  creation operators, then the second line of Eq. (\ref{eq:paulioscillators}),
  for $\tmmathbf{k}=\tmmathbf{k}'$ means
  \[ a^{\dag}_{\tmmathbf{k}} a^{\dag}_{\tmmathbf{k}} + a^{\dag}_{\tmmathbf{k}}
     a^{\dag}_{\tmmathbf{k}} = 2 a^{\dag}_{\tmmathbf{k}}
     a^{\dag}_{\tmmathbf{k}} = 0 \]
     Such a pair of operators acting on the vacuum state would return null
     result.
  In other words, if the quantum numbers are the same then we cannot expect to
  create a state with two (or more) of them, as expected from the exclusion
  principle. Thus one says that the Fermi-Dirac enumeration or statistics, applicable to $1 /
  2$-integer spin is correctly implemented using anti-commuting quantum conditions. For obtaining space-time field version of the quantisation one begins with the lagrangian density
   \[ \mathcal{L} = \overline{\psi} (i \gamma^{\mu} \partial_{\mu} - M)
     \psi \]
  where $m$ represents the fermion mass.
  This program was carried out by Jordan and Wigner. In this approach, there
  are no dangerous negative energy levels. Just as for the Klein-Gordon case, 
  the "negative energy" Dirac  solutions  can now be interpreted as
  possessing negative charge instead of negative energy, while all the
  independent solutions of Dirac equation correspond to positive energy
  particles. \

  \section{Interacting Quantum fields}\label{sec:betadecay}
  
  We saw that at the heart of Relativistic Quantum Mechanics is the possibility of
  creating and destroying quanta, and quantisation of space-time fields was developed to address this requirement. This allows construction of the free particle states of such quanta with one field introduced for every
 relevant particle species. Interaction of charged particles with photons had been taken over from Eq. \eqref{eq:minimalcoupling}.
 There still was no general prescription for the introduction of interaction among such quanta. 
However the new calculus of the creation and destruction
operators suggests how this may be done, at least for 
short range, weak interactions.

***

  {\scriptsize
  A generic way to introduce an interaction would be \cite{Dirac} through a term in the Lagrangian density that has the form
  \begin{equation}
    \tilde{\mathcal{L}}_{\tmop{Int}} = g\mathcal{V} (\tmmathbf{k}_1,
    \tmmathbf{k}_2, \tmmathbf{k}_3, \tmmathbf{k}_4) a_{\tmmathbf{k}_4}^{\dag}
    a_{\tmmathbf{k}_3}^{\dag} a_{\tmmathbf{k}_2}  b_{\tmmathbf{k}_1} 
    \label{eq:Lint}
  \end{equation}
  which we have written in the momentum space. When made to act on a state
  already containing one $a$ and one $b$ particle, it would destroy them if
  the corresponding momentum values $\tmmathbf{k}$ match, and then create fresh $a$ and $b$ particles with respective momenta of the daggered operators. The $\mathcal{V}$ which is a function of the
  four momentum values decides the energetics of this transition or scattering.
  It must necessarily contain delta-functions $\delta(k_1^0+k_2^0-k_3^0-k_4^0)\delta^3(\tmmathbf{k}_1+
    \tmmathbf{k}_2+ \tmmathbf{k}_3+ \tmmathbf{k}_4)$ with all vector momenta in-going and where $k_1^0=(|\tmmathbf{k}_1|^2+m^2)^{1/2}$ etc., ensuring the conservation of total energy and momentum.
  A crucial ingredient is the ``strength'' of such interaction, governed by
  the $g$ which called the \textit{coupling constant} . This generic construct needs to be refined as we 
  go further, and we return to it in Sec. \ref{subsec:arbitspin}
  }
  
  \subsection{Fermi's theory of Weak nuclear force}
  
  An interacting theory beyond the known electromagnetic interaction was first proposed by Enrico Fermi in 1933. It was proposed for
  understanding the beta decay of neutrons. And it was essentially along the lines of Eq. (\ref{eq:Lint}). To understand the background, recall that 
  in the early days of radioactivity, one class of radiation was called
  $\beta$ particles. These later turned out to be electrons. The so called
  $\gamma$ particles were nothing but high energy photons. The heaviest but not
  very penetrating were the $\alpha$ particles, which eventually turned out to
  be Helium nuclei with two protons and two neutrons.
  
  As we mentioned earlier, a key ingredient of nuclear matter, the neutron was ``discovered" or understood to be omnipresent rather late, by 1932. It was then understood to be the source of many known cases of beta radiation. The neutron when inside a certain class of nuclei can be indefinitely stable.
  But in a whole range of other nuclei, it is not. In fact, a free neutron is prone to decay in approximately 10
  minutes. We refer to the decay of the neutron involving emission of
  electrons as $\beta$-decay of neutrons. Fermi proposed an interaction term
  for beta decay of neutrons, which in modern form we may write as
  \begin{equation}
  \mathcal{L}_{\tmop{Fermi}} = G_F \bar{\psi}_p \gamma^{\mu} \psi_n
     \bar{\psi_e} \gamma_{\mu} \psi_{\nu}    
  \end{equation}
  The subscripts on the fields are for proton, neutron, electron and the
  neutrino respectively. When the $\psi$ fields are written out in terms of the
  creation and annihilation operators we get terms of the same form as Eq.
  (\ref{eq:Lint}), with the Fermi constant $G_F$ determining its strength.  For interpreting the interaction one may think of the
  unbarred fields as destroying an incoming particle and barred
  fields as creating the corresponding particle from the "in" or early time vacuum. If one wants to focus on their role in the "out" or late time vacuum, then the destruction of a in-vacuum particle gets traded for
  the creation of an out-vacuum anti-particle. Thus the
  term above can be read as destroying a neutrino and creating an electron and
  destroying a neutron and creating a proton. In practice, a neutron of in-vacuum 
  disappears, with the appearance of a proton, an electron and an
  anti-neutrino in the out-vacuum. Since there is no neutrino to begin with, the symbol causing {\tmem{destruction}} of neutrino equally
  well serves to {\tmem{create}} the anti-neutrino in the out-vacuum. Recall that a total energy-momentum conserving delta-function is always encoded. In the present case, the rest-mass of the
  neutron is sufficiently large to provide rest masses to the proton and the
  electron and yet leave behind some kinetic energy
  that can be carried away by the electron emerging as the $\beta$ particle
  and by the invisible, almost massless anti-neutrino.
  
  When this theory was first proposed, Fermi sent it to a very reputed journal   where it was rejected. 
  We quote this here primarily as a
  solace to all working scientists who may have found their most exciting work rejected. We know from later developments, that Fermi's proposed interaction did not have all the information that the weak
  force incorporates. But it was a bold proposal which opened up the use of quantum field theory to newer forces of nature. It was adequate for calculating the decay rate of a neutron. However its implications were as profound as
  it dealt with a new force of nature, and thus introducing a new constant of nature, the Fermi  constant (in natural units)
  \[ G_F \approx 1.17\times 10^{- 5} \tmop{GeV}^{-2} \]
The Standard Model we spoke of in the Introduction, and the Electroweak theory we mention in Sec. \ref{subsec:recapitulation} in Conclusion, have their origins in this first foray into the new force law.
  
  \section{QED }  \label{sec:qed}

  The Dirac equation permitted successful such as for the Hydrogen atom levels, at energies of about $10$eV, far lower than the rest mass energy of the electron, $m c^2 = 511$ keV. 
  The one-particle interpretation worked well in this case. 
  But it was soon emerging that there was a ``fine structure'' to the Hydrogen levels, with small
  splittings within the spectral lines. The constant
  $\alpha = e^2 /4\pi \hbar c$ which determines the size of these splittings came
  to be known as the fine structure constant.  The results of the Dirac equation gave a dependence of the energy levels only
  on the principal quantum number $n$, and the total angular momentum $\tmmathbf{j}=\tmmathbf{l}+\tmmathbf{s}$ of the level. According to this, at $n = 2,$ two particular levels, $2S_{1 / 2}$ and  $2P_{1 / 2}$ would have the exact same energy values. 

   Dirac theory could be used in many contexts successfully, provided one invoked the negative energy states to complete the Hilbert space in which the calculation was being done. Early calculations such as the Klein-Nishina formula for the electron-photon scattering called Compton scattering, and the Bhabha scattering formula for $e^+-e^-$ collisions relied on this old framework successfully. But in trying to understand electron self-energy due to interaction with photons, one got formally infinite expressions. The new framework that finally provided a systematic procedure to go beyond the simple QFT and avoided infinities has come to be called Quantum Electrodynamics, QED.

  \subsection{The Lamb shift}\label{subsec:Lambshift}
  After the second world war, microwave technology had improved substantially.
  Precise measurements of the Hydrogen energy levels showed a tiny but inescapable discrepancy, the ``Lamb shift'' in the energy levels. This  was detected in ingenious experiments of Lamb  and Retherford, reported in 1947 by 
  Lamb. The Lamb shift found a  splitting in the two levels $^2S_{1/2}$ and $^2P_{1/2}$. 
  This was about $10^9$Hz, or parts per million deviation in the main value of the transition frequency $2.5\times 10^{15}$ Hz of the Lyman "alpha" or the strongest line of Hydrogen. At this level, several other competing corrections also arise and which are larger, but the discrepancy cannot be closed unless QED processes are considered to a higher level. 
  
  \subsection{A new diagrammatic calculus}
  In theoretical calculations almost all realistic problems are unsolvable. When the interactions can be treated as a small correction to the solvable problem, a special method called perturbation theory is employed. Although this is well understood as a general strategy, in detail it can differ substantially from one class of problems to another. A perturbation theory had been in place for non-relativistic quantum mechanics. But adapting it to the relativistic case presented many subtleties.  
  
  In QED, the attempt to go to to higher orders in perturbation gave rise to uncanny singularities. The mathematical expressions set up formally, especially to incorporate the contribution of the Dirac Sea, gave infinite
  answers. Hans Bethe estimated the effect of higher order perturbation to Lamb shift, by simply cutting off the range of integration to physically relevant energies and got a reasonable answer. But the general validity of the answer was not obvious. In hindsight, one can say that the  systematic derivation of this effect needed (1) to contend with the apparent infinities, (2) and do this in way that is relativistically covariant,
  that is, any two observers in arbitrary relative motion should be able to agree on the reinterpreted answer. This is called relativistic covariance of the answer.  Truncating the range of momentum integration makes the answer depend upon the frame of reference in which that cut off was chosen. However the perturbation theory had been taken over from old
  celestial mechanics adapted to Sch{\"o}dinger's non-relativistic equation. 
  Thus relativistic covariance, even if present would not be apparent.
  
\begin{figure}[htb]
        \centering
        \includegraphics[width=\linewidth]{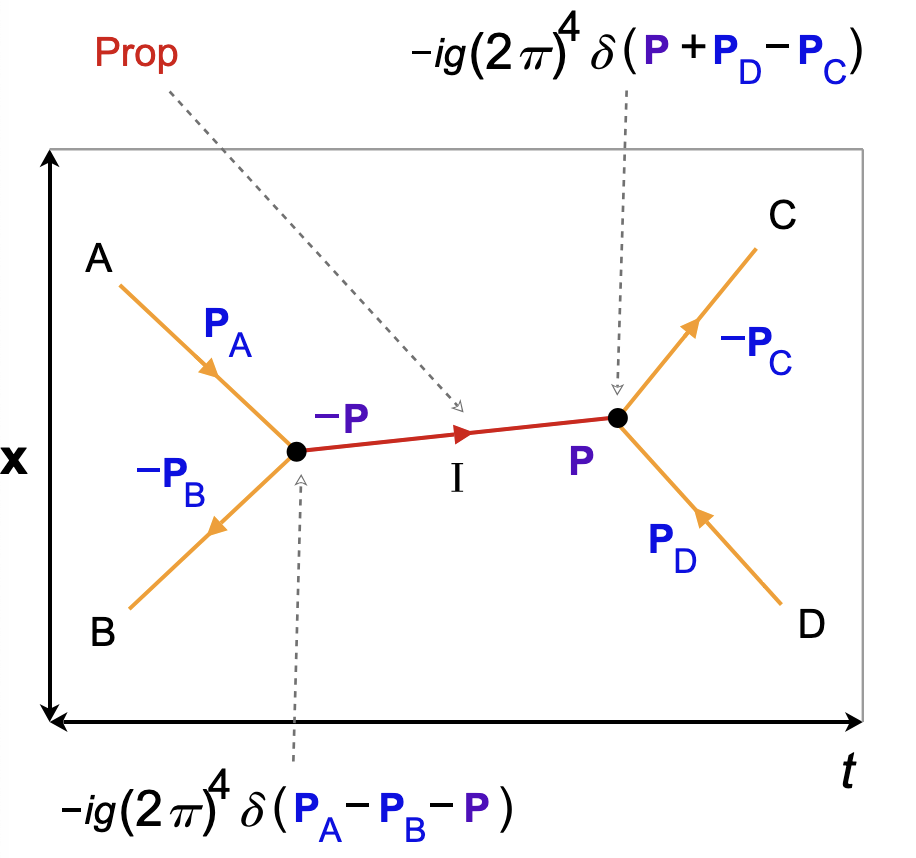}
        \caption{A Feynman diagram with due labelling suggesting the calculation to be carried out. Courtesy Wikipedia}
        \label{fig:feynmandiagram}
    \end{figure}

  A gamut of new proposals, techniques and tricks were formulated, by Pauli,
  Weisskopf, Gupta, Bleuler, Stueckelberg and others. However, they were
  consistently put together by Richard Feynman and were developed
  systematically in terms of quantum fields by Julian Schwinger. Indeed the
  calculus developed by Feynman drew upon his own insights derived from the path integral formulation of quantum mechanics which he had earlier developed as his PhD thesis at Princeton University. It was based on an earlier, little understood formulation by Dirac. This enabled him to put together the covariant Green Function
  proposed by Stueckelberg in a diagrammatic approach.  He visualised scattering processes as diagrams with lines representing propagating particles, and interactions occurring at nodes where the particle lines meet. Feynman also developed tricks that made sense of or evaluated in a usable form, integrals that would otherwise seem meaningless.
  On the other hand, Schwinger developed a systematic approach without any diagrams, with canonical methods in which generalised Green Functions formed an important ingredient. Fig. \ref{fig:feynmandiagram} shows a sample Feynman diagram.
  
  In 1947 there was a special conference held at Shelter Island.  For the first time after the war, physicists convened to focus on the problems that had been bothering them about the perturbation methods. But  importantly, they also had a salient experimental hint to proceed on, the work presented by Willis Lamb.
   Many theoretical developments happened quickly within the next year or two. Hans Bethe made his mostly correct estimate of the Lamb shift on the train back from the conference. There was a follow up conference next year at Pocono Manor in 1948.  Here both Feynman and Schwinger presented their early results. Feynman directly utilised the diagrams with lines representing "propagators" and vertices representing interaction,
  evaluated integrals magically, and got the needed answers.
    Schwinger presented his calculations concerning the same quantities. As the lore goes, the presentations of these two geniuses left the audience in  distress. Schwinger's methods were canonical but formidable.
  Feynman's methods were intuitive, but it seemed like one had to learn a  completely new language of diagrams. Once again, it was a train journey on which  Freeman J Dyson began to develop Feynman's methods systematically
  from a relativistic generalisation of established perturbation theory. He
  also showed the equivalence of Feynman's perturbation theory to the operator methods of Schwinger. Dyson completed
  this work during a subsequent visit to Princeton University. He did not have
  a doctorate degree. The Quantum Electrodynamics he developed gave him such
  a reputation that he never carried out formal doctoral work. 
  He remained a lifelong member of the Institute for Advanced Study at Princeton and continued to engage with questions of fundamentals of Physics, and of science and society,
  such as limiting the arms race, global warming and others.
  
  \subsection{Anomalous magnetic moment of the electron}
  
  Schwinger presented his methods that had found an additional divergent
  integration and had also made sense of it, to predict a departure from
  simple $g$-factor introduced in Sec \ref{subsec:gfactor}, Eq. (\ref{eq:gfactor}). The leading correction was surprisingly succinct,
  \[ \Delta g = \frac{\alpha}{2 \pi} \]
  where $\alpha = e^2 / (4 \pi \hbar c)$ is the fine structure constant, $e$ being the
  electron charge. This unexpected yet elegant formula is engraved on Schwinger's tombstone as desired by him.
  
  This modification is referred to as the {\tmem{anomalous magnetic moment}}
  of the electron.  It turns out that among many measurement methods, the
  measurement of the magnetic moment of the electron can be carried out to a
  very high precision. The current experimental value of this quantity expressed as a departure from Dirac equation prediction $g_e=2$ is,
  \[ a_{\text{e}} = \frac{g-2}{2}= 0.00115965218059 (13) \]
  with the digits in brackets representing experimental uncertainty in the last two digits.
  Schwinger's calculation was the first order correction. But there are higher
  order corrections as a Taylor expansion in the small quantity $\alpha \approx 1 /
  137$. They were pushed to a high order in $\alpha$ in a lifelong project by Toichiro Kinoshita and others. its theoretically calculated value to order $\alpha^5$ is,
  \[ a_{\text{e}} = 0.001159652181643 (764) \]
  with the last digits in brackets representing uncertainty in the
  computational answer. The two values therefore match for ten significant
  figures. This is a triumph both of the measurement techniques and of the
  computational calculus of Quantum Field Theory. As of this writing, there is a discrepancy in the magnetic moment of the muon, raising hopes of new physics to be discovered.

\subsection{Fullness of the vacuum state}
\label{subsec:vacuum}
The most important perspective to emerge from QED was the rich structure of the vacuum state $\vert \Omega\rangle$ of QFT, in a sense contradicting its very name. But one has to remember that in a theory whose interpretation always involves probabilistic outcomes, there are systematic departures from the average answer. Thus the vacuum is empty, but only on the average. If one puts systematic queries of higher order corrections in this context, one gets non-zero answers.

\begin{figure}[htb]
        \centering
        \includegraphics[width=0.8\linewidth]{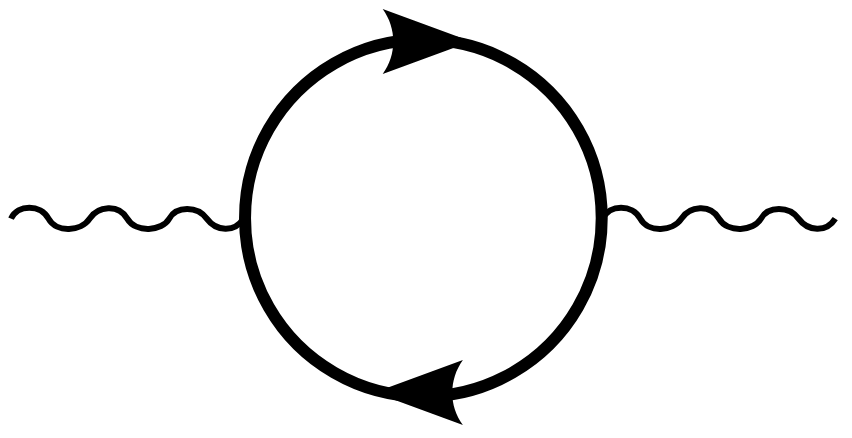}
        \caption{Visualising a spontaneous appearance and disappearance  of an electron-positron pair from a propagating photon. An effect called vacuum polarisation. The solid lines forming the loop, with arrows showing the direction of charge flow, can be thought of as an electron and a positron. Graphic courtesy Wikipedia}
        \label{fig:vacuumpolarisation}
    \end{figure}

Feynman's diagrammatic method provided a systematic method for visualising these fluctuations, or the departures from the average. The vacuum is empty and therefore charge neutral. But so is an electron-positron pair. In Dirac's hole picture they also have opposite energies, so would have the same energy as the vacuum. The vacuum could thus fluctuate to produce and reabsorb an electron-positron pair without contradicting charge or energy conservation. Such effects can be visualised as shown in Fig. \ref{fig:vacuumpolarisation}. 
The wavy lines in this diagram represent a real photon. However, the loop of solid lines which appears has to be understood as "virtual particles", that is, they cannot be observed even in principle. The diagram is only an aid to visualising and calculating.  If this "happened" within the time span permitted by the uncertainty principle there is no violation of physical laws. Indeed Schwinger's methods without any diagrams would give the same answer. This fleeting effect could show up as polarisation of the vacuum, very similar to the polarisation of neutral material under electrostatic induction due to a strong external field. In a genuine vacuum, even the polarisation effects would average to zero. But inside an atom, there is a strong electrostatic field of the nucleus, growing sharply as one gets closer to the nucleus. This means that from the electron-positron pair, the electron line would go closer to the positively charged nucleus, while the positron line would get repelled by the nucleus. This means the electron line represents a more negative contribution to the electrostatic energy than the positive contribution of the positron line. The relativistic diagrams of Feynman indeed capture not only the electrostatic but fully dynamic electromagnetic effects. This idea had been anticipated before the invention of the covariant relativistic perturbation theory, and was called the Uehling effect. Its magnitude (\cite{SW-v1}, Eq. (11.2.42))
\begin{equation}
    \Delta E= -\frac{4\alpha^5 m}{15 \pi n^3}
\end{equation}
for the $n$th level of the Hydrogen atom could be reliably calculated using renormalisation methods. It makes a small but observable contribution to the Lamb shift as mentioned earlier. 

Much has been written about the apparent non-empty nature of the QFT vacuum, even reading into it a variety of meanings of spiritual significance. Physicists should have no quarrel if some of their intellectual apparatus provides solace and edification to the lay readers. In fact, physicists are no less intrigued and obsessed with calculating and understanding various subtle effects that arise, such as light by light scattering, and the Casimir effect. But for them, the excitement lies mostly in the conceptual simplification it provides for reliable calculations, and sometimes quick estimates. 

QED vacuum is amenable to some visualisation as per these diagrams, and this is possible because the coupling constant $\alpha\approx 1/137$ is a small number. It is because the theory is weakly coupled and the interaction effects are small, that its main players can be visualised much like the free particles we see. Quantum chromodynamics (QCD) the theory of the strong nuclear force is more complicated. Its coupling constant is almost $1$ and the interaction energy dominates over the very light quarks which make up the protons and the neutrons. In fact, we know that the interaction grows ever stronger if we try to pull two quarks apart. Thus no free quarks can be seen, nor the corresponding force mediator analogues of the photons, called the gluons. Understanding and extracting more information from the QCD vacuum remains an open and exciting problem. The heavy ion collision experiments at Brookhaven and CERN among others, continue to explore QCD and its vacuum structure.

  \section{Towards modern quantum field theory} \label{sec:modernqft}
  
  Over almost three decades, Dirac's Equation continued to intrigue and tease
  physicists, challenging them for a generalisation. The  fundamental particles found till date have spin no higher than one unit of $\hbar$, and graviton though not yet detected will surely have spin $2\hbar$. No higher spin fundamental particles have been found. However, there are
  many strongly interacting particles and nuclei with spin values 3/2, 2, 5/2, 3, 7/2, {\textdots} in $\hbar$ units. Attempts continued
  for finding specific equations as well as generalised procedures to obtain
  Schr{\"o}dinger type equation obeyed by particles of higher spin. A  comprehensive solution came from a somewhat different direction as we pursue next.
  
  \subsection{Symmetry and quantum mechanics}
  \label{subsec:symmetry}
 
  One notable feature of atomic and molecular spectra was the variation in
  intensity depending on the spectral line being studied. This had to do with
  the several alternative ways the same spectral line can be generated. For
  example the molecule of methane has four Hydrogen atoms placed at the corners of
  a tetrahedron around a Carbon atom.
  
   \begin{figure}[htb]
        \centering
        \includegraphics[width=\linewidth]{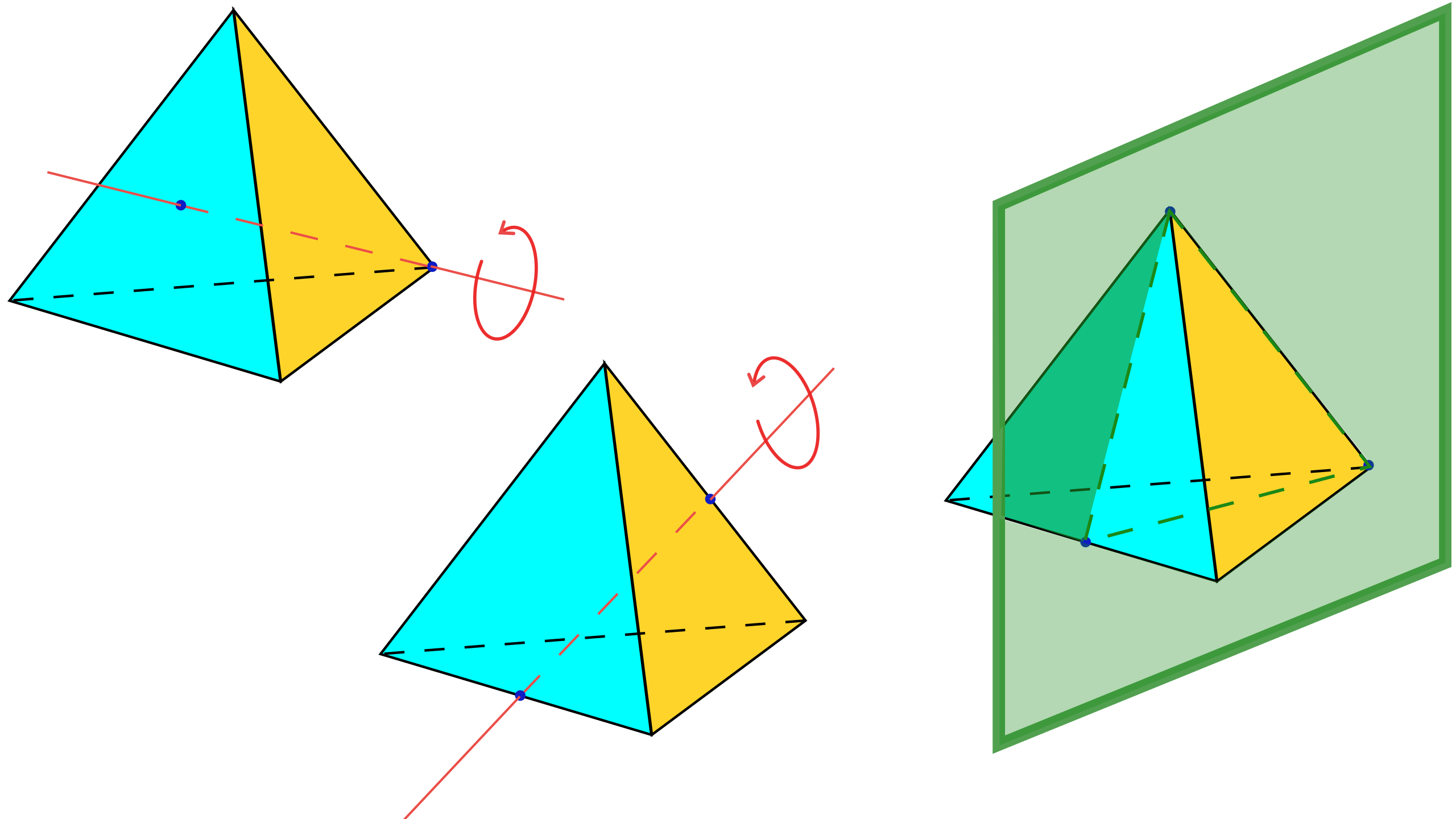}
        \caption{Possible symmetry operations on a tetrahedron. Graphic courtesy Wikipedia}
        \label{fig:tetrahedron}
    \end{figure}
  
  This geometry makes all the four Hydrogens equivalent. The corresponding
  molecular orbitals and transitions between them 
  therefore come as multiplets of identical frequencies, so called degenerate frequencies. For the methane molecule there are nine vibrational modes out which one is doubly degenerate and two are triply degenerate, and only one remains non-degenerate.   These degeneracies reflect in the relative strengths of spectral lines.
  
  From the early days of quantum mechanics, Hermann Weyl and Eugene P.
  Wigner\footnote{This author had the rare privilege of shaking hands with
  EPW on being introduced to him by John A Wheeler in his office at UT Austin, circa 1983.} had developed the mathematics needed for the use of symmetry
  groups. In mathematics itself, group theory began in the nineteenth century as a branch of algebra.
  But the developments in it had continued well into the twentieth century. Thus some of the developments in fact were carried out by Weyl,
  Wigner and others while exploring its applications to quantum mechanics. This led Wigner to write an article "The Unreasonable Effectiveness of Mathematics in the Natural Sciences" in his later years. Thus, while we cannot see the molecules, we can discern all facts about its geometry from the mathematics of symmetry. Further, Weyl and Wigner charted a course to 
  the mathematics of spin, which despite its analogy to rotations remains more subtle than rotations as we discusse earlier.
  
  It is also possible to associate symmetries with space-time itself. 
  We know that in solving physics problems, the origin of the coordinates and the orientation the axes should not affect the final answer. In this sense shifting the origin in space and time -- called ``translations'' -- as well as rotations of
  the frames of reference should not affect physics. Likewise, as per the 
  early wisdom of Galileo the relative velocity of frames of reference should also 
  not affect physics. Special Relativity added the insight that this latter principle is true, but the speed of light has to remain 
  the same in all inertial frames of reference. The symmetry group of the space-time under rotations and velocity ``boosts'' is called Lorentz Group. When space-time translations are considered in addition, the larger group is called the Poincar{\'e} Group.
  
  Interestingly, every group of possible symmetry operations could be
  ``realised'' through many different "representations''. For example, complex numbers have the structure of a symmetry group. We know
  that the complex unit $i$ obeys the rule $i^2 = - 1$. Now we can write
  complex numbers as $a + i b$, with real numbers $a$ and $b$, and carry out their algebra remembering the basic rule for $i$. 
  Now consider the 2x2 matrix
  \[ \mathcal{J}= \left(\begin{array}{cc}
       0 & - 1\\
       - 1 & 0
     \end{array}\right) \]
  Note that this matrix obeys the same rule $\mathcal{J}^2 = - I$ where $I$ is the identity matrix. We may
  therefore define a set of 2x2 matrices of the form $a +\mathcal{J}b$. Under
  rules of matrix algebra these will be just another copy of the usual complex
  numbers. These are two different representations of difference sizes for the complex numbers.
  There are more sophisticated ways in which the basic algebra of a symmetry
  group can be realised in different ways. One way to think is that the
  same symmetry operations are realised using matrices of different sizes.
  
  Moving to the core of our discussion, the 4-component Dirac wave function for a spin-$1/2$ particle of a specific mass is one of 
  the many representations of the Poincar{\'e} group. Hermann Weyl pointed out 2-component representations of spin-$1/2$ which would have to manifest as massless particles. In particular,
  E. P. Wigner worked out in detail a general procedure for obtaining
  representations of the Poincar{\'e} Group that would have any desired 
  value of space-time spin, with zero or non-zero mass.
  
  \subsection{``Feynman rules for arbitrary spin''}
  \label{subsec:arbitspin}
  
  There was however one difficulty. It was not suggested through Wigner's work
  what should be the ``wave equations'' satisfied by these larger matrix
  multiplets representing the Poincar{\'e} Group. In a series of papers in the 
  \textsl{Physical Review} in the mid-1960's Steven Weinberg elucidated the program of  
  Quantum Field Theory, armed with Wigner's methods, but also armed with several important 
  facts that were prevailing as folk theorems but not  formalised. A comprehensive
  monograph version of these papers forms the core of the famous book \cite{SW-v1}.
  
  In the approach Weinberg elucidated, it was not necessary to know
  explicitly the wave equation to be satisfied by a particle of particular
  mass and spin. What was required was the complete set of basic plane waves
  required for the description of the particle. In some sense this is Fourier analysis combined with group structure. Such plane waves 
  could be
  explicitly obtained using the methods of Wigner, simply based on the knowledge of the spin value of such particles, with non-zero or zero mass. This
  obviated first guessing an equation and then solving it to obtain those plane waves as ``solutions''.
  
  Several elegant properties of the quantum field theory approach had been
  noticed already. These were (i) locality and
  (ii) cluster decomposition, in addition to (iii) relativistic causality. Weinberg used these in reverse as criteria 
  to show that they can be ensured in scattering processes, provided 
  the theory was  constructed out of Lagrangian density made up 
  of space-time fields $\Phi (\tmmathbf{x}, t)$ which in turn had to be made up in a particular way out
  of the plane waves $u(\tmmathbf{k},\sigma,n)$ of positive energy and $v(\tmmathbf{k},\sigma,n)$ of negative energy, of Wigner's prescription. 
  Here $\tmmathbf{k}$ represents the wave number, being the momentum divided 
  by $\hbar$, $\sigma$ stands for the spin, and $n$ specifies charges that define 
  the particle species. These sets of solutions are also referred to as \textit{mode functions}. Thus, no free wave equation
  had to be solved, but the Feynman propagator could be explicitly constructed
  for any field using the representation theory of Wigner.
  
  ***

{\scriptsize 
  A lesson from the development of Quantum Electrodynamics was the Feynman
  propagator as an essential building block for the systematic corrections to
  motion of particles due to interaction with other particles. The propagator is
  pictorially a line in a graph, but for using it, one has to associate a
  mathematical expression to the line. Feynman had
  successfully used propagators for photons and electrons which for the present purpose can be written in the forms
  \begin{eqnarray*}
    G_{\tmop{photon}} & = & \frac{k_{\mu} k_{\nu} - \eta_{\mu \nu} k^{\rho}
    k_{\rho}}{k^{\rho} k_{\rho} + i \varepsilon}\\
    D_{\tmop{electron}} & = & \frac{\gamma^{\mu} p_{\mu} + m}{p^{\rho}
    p_{\rho} - m^2 + i \varepsilon}
    \label{eq:ph-el-propagators}
  \end{eqnarray*}
  In a series of papers outlining the construction of Quantum Field Theory
  based on the Poincar{\'e} Group and the single particle mode functions, Steven Weinberg elucidated the program of Quantum Field Theory in
  the following steps

  \begin{enumerate}
    \item Each particle species of a specific mass and spin is a particular representation worked out by Wigner.
    
    \item In the spirit of Fock and Dirac we introduce creation operators
    $a^{\dag}_{\tmmathbf{k}, s}$ and destruction \ operators $a_{\tmmathbf{k}, s}$ for particles represented by positive energy modes, and creation operators $b^{\dag}_{\tmmathbf{k}, s}$ and
    destruction \ operators $b_{\tmmathbf{k}, s}$ for anti-particles represented by negative energy modes. Here
    the subscript $\tmmathbf{k}, s$ refers to the momentum or wave
    number $\tmmathbf{k}$ of the specific mode and $s$ the spin projection of that mode along a referred axis.
    
    \item We need to introduce the particular field expansion,
    \begin{eqnarray*}
      \Phi (\tmmathbf{x}, t) &\sim& 
      \sum_{\tmmathbf{k}, s} \{ a_{\tmmathbf{k}, s} u_{\tmmathbf{k}, s}
      (\tmmathbf{x}, t) e^{- i (\omega_{\tmmathbf{k}} t -\tmmathbf{k} \cdot
      \tmmathbf{x})} \nobracket  \\
     \nobracket& & +\,  b^{\dag}_{\tmmathbf{k}, s} v_{\tmmathbf{k}, s}
       (\tmmathbf{x}, t) e^{+ i (\omega_{\tmmathbf{k}} t -\tmmathbf{k} \cdot
       \tmmathbf{x})} \} 
        \end{eqnarray*}
    These $u$ and $v$ are the mode functions from Wigner's representation theory for each label $\tmmathbf{k}, s$. Here $\omega_{\tmmathbf{k}}$ $=\sqrt{|\tmmathbf{k}|^2+m^2}$. The hermitian conjugate field $\Phi^{\dag}$ is also introduced, with complex conjugate mode functions.

    \item The key construction is the numerators occurring in Feynman propagators such as $G$ and $D$ of Eq. \eqref{eq:ph-el-propagators}. These are special matrices, ``spin sums'' to be constructed out 
    of the $u$ and $v$ functions. In the language of linear vector spaces, these are closely related to the completeness relations for the mode functions.
   
      \item The denominator remains the same for all particles, with the
      corresponding value of the mass which can be zero or non-zero. This is
      because all that the denominator does is to remember the relativistic energy-momentum
      relation $(k^0)^2 +\tmmathbf{k}^2 - m^2 = 0$, regardless of the spin of the particle. It introduces a pole at the location $k^0= \omega_{\tmmathbf{k}}$. When the propagator occurs as an internal line in a loop, all the four variables, $k^0$ and $\tmmathbf{k}$ are to be freely integrated, not restricted by the energy-momentum relation. This makes these particles unphysical, and called "virtual".
 
    \item Given a theory of any set of particles of specific masses and spins, the interaction vertex in the Feynman diagram will be decided by the Lorentz scalar lagrangian density that can be permitted by the symmetries of the fields.
    
  \end{enumerate}
  }

  \subsection{QFT of Massless particles}\label{subsec:massless}
  
  As students of quantum mechanics know, a particle of spin $J$ 
  occurs in $2J+ 1$ projection states, with the $z$-component of angular momentum ranging
  from $- J$ in integer steps to $+ J$, in the units of ${\hbar}$. In Wigner's
  analysis, massive relativistic particles offer a simplification. We can first go to their
  rest frame and classify all the possible states and then simply apply a
  velocity ``boost'' transformation bringing them to the required momentum. In
  the rest frame the states are the same as those allowed by non-relativistic
  quantum mechanics due to rotational symmetry, the same as the $2J+1$ states. 
  Thus the Lorentz covariant mode functions can be explicitly constructed.
  
  However, massless particles are special. They are always moving at the speed
  of light. We cannot go to their rest-frame. More correctly, there is no
  transformation available to boost ourselves to the frame of light, because
  the parameter $1 / \sqrt{(1 - v^2 / c^2)}$ needed in the boost
  transformation diverges. A detailed analysis shows that massless particles
  have only two projection states, $- J$ and $+ J$. For example 
  the photon having ``spin-1'' does not have 3 possible projection states $- 1, 0, + 1$.
  It has only two states, $+ 1$ and $- 1$. These states can be though of as,
  respectively, the photon spinning clockwise looking down its momentum vector
  or anti-clockwise looking down its momentum vector, but only with its
  maximum allowed value $J$. Classically the two states of photons find
  manifestation as two types of polarisation of electromagnetic waves.  This fact is
  generally true for all massless particles. We therefore refer to their spin
  states as ``helicity'' rather than spin. The word spin is preferably used
  only for massive particles whose spin states can be counted 
  as $- J \ldots +J$ in their rest frame using the usual angular momentum algebra.
  
  Weinberg also obtained new fundamental facts about massless particles. For
  consistency of QFT in producing unitary $S$-matrix, the massless particles
  can only couple to conserved currents. Thus photons of helicity $1$ must
  couple to a conserved rank $1$ tensor, or a vector current. This ensures the 
  well known fact that the electromagnetic charge is conserved.
  In classical electromagnetism, current conservation is related to
  gauge invariance of electromagnetic potentials. 
  Now the same relationship is seen to emerge in a different language
  from the more fundamental relationship of massless quanta and 
  unitarity of quantum field theory. Likewise, the as yet
  undiscovered graviton which must have helicity $2$, must couple 
  only to a conserved symmetric tensor of rank $2$. The obvious, and 
  in our phenomenological experience unique candidate for such a conserved rank $2$ tensor is the energy-momentum tensor. 
  Now there is a theorem known in General Relativity proved by Suraj Gupta and others, that a linear theory of second rank tensor field, if consistently coupled to a conserved current, must reproduce the full non-linear equations of General relativity. And Weinberg's result now requires that
  a spin-2 massless must couple to a conserved second rank current. Thus General  Relativity appears to be the inevitable theory resulting from the existence only of massless spin-2 particles and the consistency of the scattering matrix. An accessible exposition of this is given in \cite{{Mukhi-on-SW}}.
  
  \section{``$P C T$, Spin and Statistics, and all that''}\label{sec:CPT}
  
  That is the title of an erudite book,  a monograph concerning an  axiomatic approach to quantum field theory by its pioneers, Raymond F. Streater and Arthur S. Weightman. Let us briefly explain the topics referred to in this title.
  
  \subsection{$C P T$}
  
  $C P T$ is the more common order in which these letters are written. The physics of fundamental particles kept throwing up one uncanny fact after
  another through cosmic ray and terrestrial experiments. As we saw earlier,
  anti-particles originally gleaned from Dirac equation came to be accepted
  after the discovery of positrons. Likewise there were anti-muons, and later
  anti-protons and so on. And particles and anti-particles were always of
  identical masses but opposite values of all quantum numbers including the
  electric and other interaction charges. This came to be called Charge
  Conjugation symmetry of the world, represented by the symbol $C$.
  
  Just when people began to expect that there was always going to be an
  anti-particle due to opposite signs of energy being on the same footing in relativistic 
  quantum mechanics, there came a big surprise. It was discovered in 1956,
  that the neutrino has no anti-particle. But the neutrino was also found to
  be massless (this understanding has changed after the 1990s). The anti-neutrinos  produced in the decay of the neutrons were always right-handed. It was then  realised that in the decay of the anti-neutron, there would occur only the neutrinos, but they would all be left-handed. \
  
  Here right-handed means that the sense of spinning as the neutrino goes
  forward is same as the sense in which a screw moves as 
  it is being turned, and if the sense of spin is anti-parallel to the direction of motion we call the
  particle left-handed. Now if we hold a mirror to the emerging neutrinos in an experiment, the
  direction of motion would look flipped. But the sense of rotation does not change. The reader is invited to check this by holding a mirror. Thus the
  mirror image of the left-handed neutrino would be a right-handed neutrino.
  But no such particle was found in nature, not in the weak nuclear processes.
  Likewise no left-handed anti-neutrino has been found in weak nuclear interactions.
  
  This operation, of comparing a particle with its mirror image is called checking for parity of the particle, represented by the symbol $P$. In physics we know that a vector changes sign under parity, while a pseudo-vector does not change sign under parity. \textit{Alice in Wonderland} is written
  on the premise that the mirror world can be a source of bizarre occurrences.
  In the case of neutrino, the mirror image was found to not exist in nature! It was found
  that the neutrino emerging in weak interactions was purely left handed. And its mirror image never appeared in these experiments. On the other hand, a right handed anti-neutrino indeed existed, but no left handed anti-neutrino. Thus for the neutrinos occurring in  weak interactions, neither $P$ nor $C$ were valid symmetries, but the
  combination $C P$ was. The proposal that parity may not be conserved was made by Chen-Ning Yang and Tsung-Dao Lee. It was experimentally verified by Chien-Shiung Wu 
  in 1956. By 1957 E. C. George Sudarshan and Robert E. Marshak presented the hypothesis that this was a universal feature of the weak nuclear force, cryptically called 
  the "V - A" theory. It is this structure of the interaction Hamiltonian and absence of the V+A structure that results in only the left handed neutrino and its $C P$ conjugate, the right -handed anti-neutrino being involved in the weak interaction. 
  
  Soon after, in 1964,  a remarkable experiment was carried out by 
  Val Fitch and A. J. Cronin. They chose to study the newly discovered 
  heavy particles called $K$-mesons further. These were a completely new form of matter, and
  contained what came to be called the ``strange'' quark. To everyone's
  surprise, it was found that even the combined symmetry $C P$ was not
  conserved in the decay of $K$-mesons. However, there was one
  reprieve. Like flipping space in the mirror, one could also consider
  flipping the arrow of time, at least as a thought experiment. This is like
  running a movie backwards, and the operation is denoted $T$. It was found that if time reversal symmetry $T$
  was invoked, the overall combination $C P T$ was indeed a symmetry of the  $K$-meson system.
  
  Interestingly, as mentioned in Sec. \ref{subsec:symmetry}, Weyl had already considered a version of the Dirac equation which would be satisfied by a species that could be purely left handed or purely right handed. However such particles would have to be massless. Indeed, any attempts at measuring the neutrino mass directly in weak interaction have returned negative results to date, which tallies well with it being only one kind of Weyl species. Over the decades, however, this picture has changed. At least two of the three known species of neutrino are estimated to have masses, though extremely small,  $0.1$-$1$ eV. As pointed out in Sec. \ref{sec:peculiarities}, the three species oscillate or spontaneously transmute into each other. We will be unable to describe this here any further. However we mention a further generalisation for spin-$1/2$ fermions which was proposed by the elusive Italian physicist 
  Ettor{'e} Majorana. He showed that it was possible to have massive spin-$1/2$ particles which are their own charge conjugates. Such particles will carry no charges, and very likely make up some part of the transmuting triplet of neutrinos, which now carry some mass.
  
 Majorana neutrino is one more possible representation of the Lorentz group. An intense search is on for two decades at Karlesruhe to decide whether the standard neutrino "mixes" with such a species. Interestingly, these facts have a significant bearing on the matter-anti-matter asymmetry of the Universe. It is believed, as originally proposed by Andrei Sakharaov in 1967, that $C$ and $CP$ violating interactions are required to obtain the matter-anti-matter asymmetry of the Universe. Further, for the $C P$ asymmetry to play out, we must have a definite arrow of time. If Majorana neutrinos exist, then the hot big bang universe provides all the necessary conditions for a dynamical explanation of the matter-anti-matter asymmetry of the Universe. These are subjects of extensive current study.  
 
 Weyl and Majorana type excitations are now commonly invoked in condensed matter systems. Thus the classification of particles provided by relativistic quantum mechanics is also suggesting exciting possibilities in quantum computing applications.
  
  \subsection{Spin and statistics}
  
  The word ``statistics'' is misleading for the context in which we are going
  to discuss it here. Statistics suggests some averaging, some lack of
  information and so on. Here, statistics is not involved in that sense at
  all. Historically some fundamental rules of quantum mechanics were deduced
  through thermodynamic states of the quanta. This framework is called statistical mechanics, and hence the descriptive term "statistics" continues to be applied to these new fundamental  
  rules of enumeration, which have nothing statistical about them.
  
  We have earlier seen how bosons and fermions are quantised. It had been clear
  from atomic physics and hence Pauli exclusion principle, that electrons
  which carried spin $\hbar / 2$ had to be quantised via anti-commutators.
  Similarly, Bose's derivation of Planck spectrum had made it clear that
  photons of spin $\hbar$ had to be quantised via commutators. From
  considerations of relativistic quantum field theory it became 
  clear that all
  the particles carrying spin in half-integer units of $\hbar$ should be
  quantised via anti-commutators and all particles carrying spin in integer
  multiples of $\hbar$ should be quantised via commutators.
  
  Over the decades, many versions of the proof that this is how the fields
  need to be quantised emerged, especially also in the formal book whose title is the  title of this main section. In Weinbeg's treatment, this differentiation is
  required to ensure the relativistic causality of the S-matrix. The theorem
  remains a cornerstone of the grand developments in the theory of quantised fields.
  
  \section{Conclusion}
  
  \subsection{Accepting the quantum}
  
  Heisenberg's seminal 1925 paper that effectively launched quantum mechanics \cite{Heisenber1925}\cite{Aitchison:2004cic} 
  begins with the most radical, if preposterous proposal. It calls for
  `discard[ing] all hope of observing hitherto unobservable quantities, such
  as the position and period of the electron'. Heisenberg's approach, and
  quantum mechanics in general, continue to baffle the succeeding generations.
  Yet, in hindsight we know that Heisenberg's point of view has to be
  accepted, even if grudgingly. We are living with such enigmas arising from
  theoretical physics already, just that we are resigned to them. For example, we accept the idea of electromagnetic "waves" propagating
  through {\tmem{nothing}}.  And everybody accepts these concepts
  without any grudge. The coming generation is going to see devices and computers  
  based on quantum mechanics in their daily lives, and should be freed from the prejudices of the previous generations. We must therefore welcome quantum mechanics, marveling, as Dirac suggests, on its linear superposition principle rather than complain about its probabilistic prediction framework. In any case, the idea of instantaneous velocity was only an idealisation, and so is the continuum, so microscopic physics showing departures from these assumptions need not surprise us.
  
  \subsection{Recapitulation}\label{subsec:recapitulation}
  
  The lesson of early attempts at discovering relativistic ``wave equations''
  was that their interpretation as describing Schr{\"o}dinger wave
  functions was problematic, not least because the Schr{\"o}dinger approach
  fixes the total number and type of particles by normalising the wave
  function. Relativistic phenomena clearly indicate that quanta can be created
  and destroyed and inter-converted in interactions. At first this knowledge
  was restricted to photons which were emitted or absorbed by atoms and
  molecules. But the experience with cosmic rays followed by the accelerator
  experiments showed this to be a general feature.
  
  As such, Quantum Field Theory becomes the required
  framework. Further, the information about spin is correctly accounted for if we identify the correct representation the particle belongs to under the Lorentz group. In the older literature this is referred to as ``second''
  quantisation, but for relativistic particles this is the ``first'' and only quantisation. Weinberg's works also showed that QFT was in fact the minimal yet most
  comprehensive framework ensuring the principles of unitarity, locality and relativistic causality.
  
  To a substantial extent the framework of quantum field theory is geared towards perturbation
  theory, using the language of the $S$-matrix. It can be used for any kind of quanta whose rest-mass, spin and  conserved charges are known, and are weakly coupled. Renormalisable  quantum field theory is the
  de facto framework for analysing the experiments at all the high energy physics experiments, including the Large Hadron  Collider. This brings us to a bedrock principle for both the weak and the strong force, the gauge principle. This is a generalisation of the gauge principle encountered in electromagnetism. And renormalisation has been successfully
  applied to the eletroweak gauge theory of Glashow, Salam and Weinberg, which turns out to be weakly coupled.   For quarks and hadrons interacting
  via the strong nuclear force, the field theory approach can still be used to write out the basic theory,
  the action function, using just the properties of spin and the conserved
  charges. The theory also works perturbatively at energies higher than about 1 GeV. But the perturbation approach does not work at lower energies.  
  Yet the gauge principle combined with field theory provides very useful insights. 
  
  Quantum Field Theory has sometimes demanded specific new particles  for the consistency of the known ones. To give an example, a new species of
  quark the charm quark was predicted in 1974 based simply on the information of lower energy particles and the principles of relativistic quantum theory.
  Likewise, the theory of the Higgs boson is a beautiful example. Proposed by
  Weinberg and Salam in 1967-68, it spurred the search for new heavy particles the $W^{\pm}$ and the $Z^0$ bosons which were discovered in the 1980's. The Higgs  particle provided a framework for understanding masses of all the particles in the electroweak sector within quantum field theory. Hence the
  search for this one particle was considered one of the major justifications
  for constructing the ambitious Large Hadron Collider. This heavy scalar  particle was indeed discovered in 2012, 45 years after its prediction. 
  
  Further, given a list of particles and their masses and spins, and their
  conserved charges, quantum field theory substantially restricts the kind of
  mutual interaction they can have. This is an exercise called ``model
  building''. For example, as of now Dark Matter is known only through its
  presence at cosmic scales. But many specific models for it can be proposed according to these rules, and then the model can be counter-checked against existing and emerging data. These are studies in progress.
  
  Having got this far, we may well say, ``and for everything else, there is
  string theory'', or something else like it. We can say quantum field theory is the minimal framework
  that forms the basis for all the relativistic quantum phenomena. However, let us recall the great 
  dichotomy in mathematical natural science. There is always the framework of mechanics, and then there are force laws. 
  The framework does restrict the kind of force laws one can have to some extent, but not substantially. The program started by Newton has continued through the Lagrangian and Hamiltonian developments.
  With Einstein's geometric vision for General Relativity, there was a possibility that all of physics will
  seamlessly emerge as geometry, with no dichotomy of mechanics framework versus a force law. 
  But the other three fundamental forces have failed to fit into this paradigm, especially since at least the strong and weak force manifest only at quantum level. String theory seems to fulfil the aspiration
  of uniting the framework of mechanics with the prescription of force laws. It predicts all the possible particles with their masses, spins, charges as well as all their possible interactions. In the list of predicted particles, it also
  includes the graviton naturally. As we saw in Sec. \ref{subsec:massless}, that also means General Relativity is included in this theory, including its elusive microscopic version, a theory of quantum gravity. What is more, it also allows the determination of the
  number of space-time dimensions in which the theory can consistently unfold, viz., 9 space and 1 time dimension. Unfortunately it
  can manifest at low energy in myriads of ways, and it is not clear how it may be reconciled with the physics we have. A few specific predictions it can
  make about the world are incorrect. So perhaps there is some other approach, or there is still a guiding principle within string theory that will help resolve these issues. Whatever may be the answer to that, we believe that at the energy scales of experience, physics will manifest as a set of particles and fields carrying gauge charges and obeying Poincar{\'e} symmetries. So QFT  is going to remain a durable  framework for foreseeable future. 
  
 The corresponding author states that there is no conflict of interest. 

\section{About the author}
\begin{figure}[htb]
        \centering
        \includegraphics[width=0.4\linewidth]{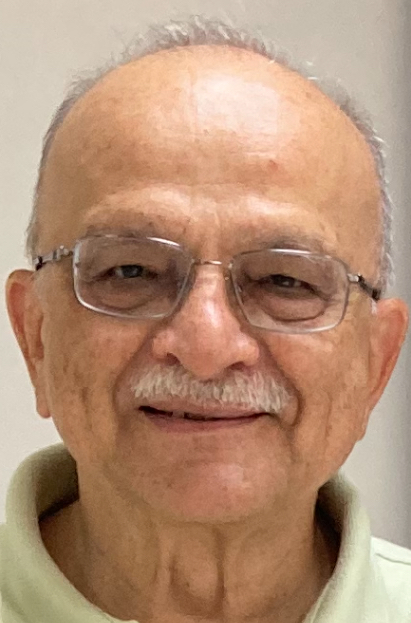}
    
        \end{figure}

U. A. Yajnik obtained his integrated MSc Physics from IIT Bombay and PhD in Elementary Particle Physics from the University of Texas at Austin. His thesis advisor was E C G Sudarshan. During this time he had the opportunity to attend four semesters of Quantum Fields Theory courses offered by Steven Weinberg, based on the first two volumes of the latter’s well known eponymous textbooks. Subsequently he was a Postdoctoral Associate with S. Weinberg. On returning to India he was a visiting Fellow at TIFR Mumbai. He joined the faculty of IIT Bombay in 1989, serving there till superannuation, intermittently holding Chair Professor positions and Dean positions. As of 2023 he is a Visiting Professor and Faculty In-charge Dean of Faculty Affairs at IIT Gandhinagar.



\begin{thebibliography}{99}
  
  \bibitem{SW-v1} Steven Weinberg, \textsl{Quantum Theory of Fields, Volume I, Foundations}, Cambridge University Press, 1996.
  
  \bibitem{IandZ} C. Itzykson and J-B Zuber, McGraw Hill Pub. Co., 1980  

\bibitem{Gottfried:2010kn}
K.~Gottfried,
``P.A.M. Dirac and the Discovery of Quantum Mechanics,''
Am. J. Phys. \textbf{79} (2011), 261
doi:10.1119/1.3536639
[arXiv:1006.4610 [physics.hist-ph]]

\bibitem{Gingrich} Douglas M. Gingrich, \textsl{Practical Quantum Electrodynamics}, CRC Press, 2006

\bibitem{Born:1926uzf}
M.~Born, W.~Heisenberg and P.~Jordan,
Z. Phys. \textbf{35} (1926) no.8-9, 557-615
doi:10.1007/BF01379806; paper 15 in B.~L.~van~der~Waerden, ``\textit{Sources of Quantum Mechanics}" North Holland Pub. Co. 1967

  \bibitem{Mukhi-on-SW} Sunil Mukhi, \textsl{Resonance} Volume \textbf{28} Issue 6 June 2023 pp 985-998
  
  \bibitem{Dirac} P. A. M. Dirac \textsl{Principles of Quantum Mechanics}, 4th Ed. (revised), Oxford University Press, 1958. See Chapter X.

\bibitem{Heisenber1925} W. Heisenberg \textsl{Z. Phys.} \textbf{33}, 879 (1925); Paper 12 in \textit{op. cit.}, \cite{Born:1926uzf}

\bibitem{Aitchison:2004cic}
I.~J.~R.~Aitchison, D.~A.~MacManus and T.~M.~Snyder,
Am. J. Phys. \textbf{72}, no.11, 1370-1379 (2004)
[arXiv:quant-ph/0404009] provides a modern paraphrasing of the Heisenberg paper above.
  
  \end{thebibliography}
\end{document}